\documentclass[aps,notitlepage,nofootinbib]{revtex4-1}

\usepackage{bm,amsmath,amssymb,graphicx,MnSymbol}

\usepackage{footnote}

\usepackage{enumitem}

\newcommand{\bd}{{\bm d}}
\newcommand{\bL}{{\bm L}}
\newcommand{\bom}{{\bm \omega}}
\newcommand{\bM}{{\bm M}}
\newcommand{\bMh}{{\bm {\hat M}}}
\newcommand{\bH}{{\bm H}}
\newcommand{\bP}{{\bm P}}
\newcommand{\bv}{{\bm v}}

\newcommand{\bI}{{\bf I}}
\newcommand{\bK}{{\bf K}}
\newcommand{\bT}{{\bf T}}

\begin{document}

\title{An integrable family of torqued, damped, rigid rotors}
\author{J. A. Hanna}
\email{jhanna@unr.edu}
\affiliation{Mechanical Engineering, University of Nevada, 1664 N. Virginia St. (0312), Reno, NV 89557-0312, U.S.A.}

\date{\today}

\begin{abstract}
	Expositions of the Euler equations for the rotation of a rigid body often invoke the idea of a specially damped system whose energy dissipates while its angular momentum magnitude is conserved in the body frame.
	An attempt to explicitly construct such a damping function leads to a more general, but still integrable, system of cubic equations whose trajectories are confined to nested sets of quadric surfaces in angular momentum space.  For some choices of parameters, the lines of fixed points along both the largest and smallest moment of inertia axes can be simultaneously attracting.  Limiting cases are those that conserve either the energy or the magnitude of the angular momentum. 
	Parallels with rod mechanics, micromagnetics, and particles with effective mass are briefly discussed.
\end{abstract}

\maketitle

A classic pedagogical example \cite{Arnold89, ThompsonStewart02} in integrable nonlinear dynamics is that of the Euler equations describing rigid body rotation in terms of the angular momentum in a body-fixed frame.
The undamped equations conserve two quantities: the magnitude of the angular momentum and the rotational kinetic energy.  These can be visualized elegantly in angular momentum space as a series of nested spheres and ellipsoids, respectively, with trajectories of the system corresponding to intersections of these families of surfaces.
The effect of damping is introduced qualitatively, by asking the student to consider a process that reduces the energy while preserving the magnitude of the angular momentum.  A picture is provided of a spiral trajectory taking the body from a maximum to a minimum energy point on an angular momentum sphere, that is, from rotating around its smallest to its largest moment of inertia axis.  This spiral is, presumably, the result of a global bifurcation destroying the structurally unstable elliptic points and heteroclinic trajectories between saddle points of the conservative system.
The mechanism for this process is left unspecified, but the example of the Explorer 1 satellite may be invoked \cite{BracewellGarriott58, KrechetnikovMarsden07}, with the implication that perhaps the body is slightly deformable and is spinning in a vacuum, not subject to external torques.

This note arises from an attempt to explicitly construct a damping term that leads to the behavior of this well known, contrived example.
I find that a simple construction provides both this term and another with reciprocal behavior, a momentum-reducing term that conserves the total energy.  A combined dynamics featuring both terms can stabilize one or both of the largest and smallest moment of inertia axes, and still preserves a constant of the motion that restricts trajectories to families of quadrics in momentum space, which allows us to determine the unique pair of attractors corresponding to given initial conditions.
Upon further reflection, it appears that this simple example illustrates the construction of special types of conservative and dissipative terms that have been considered in more formal contexts involving bracket formulations of mechanics.
I conclude with brief suggestions of where similar dynamics might be found in, or applied to, other physical systems.

\section{The dynamics}

An asymmetric rigid body has a moment of inertia tensor $\bI$ with distinct principal values $I_1 > I_2 > I_3 > 0$ and constant entries in an attached frame.  In this frame, the body's angular velocity $\bom$ and angular momentum $\bL$ are related by $\bL = \bI \cdot \bom$, and $\dot\bL\cdot\bom = \bL\cdot\dot\bom$. 
Consider the equations of motion
\begin{align}
	\dot\bL = \bL \times \bom + \bd \, , \label{eqofmotion}
\end{align}
which are those of a freely rotating rigid body, augmented with a damping function $\bd$. Without damping, the energy $H = \frac{1}{2}\bL\cdot\bom$ and squared angular momentum $L^2 = \bL\cdot\bL$ are conserved.
Because $\bI$ is positive definite, both quantities are non-negative.

In the presence of damping, these quantities change as follows:
\begin{align}
	\dot H = \dot\bL\cdot\bom = \bd\cdot\bom \label{Hdot} \, , \\ 
	L\dot L= \dot\bL\cdot\bL = \bd\cdot\bL \label{Ldot} \, .
\end{align}
If we seek a function $\bd$ such that \eqref{Hdot} is negative while \eqref{Ldot} is zero, we could write the equations \eqref{eqofmotion} in component form and construct many different families of expressions with permutation symmetry in the components that satisfy these requirements.  However, such terms are unlikely to be physically meaningful.  
Instead, let us presume that no additional external forces or fields are present in the problem, and construct (pseudo)vector 
 functions of increasing order in angular velocity using the available physical objects $\bom$, $\bL$, and $\bI$, and natural operations such as the dot and cross product.  Let us also assume that all of the relevant geometry of the body is encapsulated in $\bI$; this is in contrast to a situation such as rolling, where symmetry is broken and a geometric vector could appear.  
Finally, let us assume that the only qualitatively important contribution of the moment of inertia tensor $\bI$ will be through its appearance in the angular momentum $\bL = \bI\cdot\bom$.  This means that we will not consider quantities such as $\bI\cdot\bI\cdot\bom$.  Any use of invariants of $\bI$ could be absorbed into constant coefficients, and so is not an important complication.

With these restrictions, the only linear vector terms are $\bL$ and $\bom$.  Either of these terms alone will damp the system to zero angular velocity.  Neither alone nor in linear combination can these terms allow \eqref{Ldot} to be zero, so long as $\bL$ and $\bom$ are not collinear or orthogonal, which possibilities are prevented by the the properties of our asymmetric, positive definite $\bI$, except when the system aligns with a principal axis.
The same observations can be made about \eqref{Hdot}. 
The only quadratic vector term is $\bL\times\bom$, which is the type of term we already have.  This always conserves both quantities $H$ and $L^2$.
Therefore, the lowest order terms to consider are cubic.  Of the eight possible terms, six of them, namely (defining $\omega^2 = \bom\cdot\bom$) $\omega^2\bom$, $2H\bom$, $L^2\bom$, $\omega^2\bL$, $2H\bL$, and $L^2\bL$, behave similarly to the linear terms. 
 However, the remaining two terms are of interest.  We consider
\begin{align}
	\bd =  \epsilon_H \bom\times\bL\times\bL + \epsilon_L \bL\times\bom\times\bom \, , \label{damping}
\end{align}
where the coefficients $\epsilon_H$ and $\epsilon_L$ are positive constants with units of inverse angular momentum and inverse angular velocity (time).  The first term acts orthogonally to the angular momentum and the second to the angular velocity.
It is clear from inspection that when $\epsilon_L = 0$, $\bd\cdot\bL = 0$ and $\bd\cdot\bom \le 0$, so the first term has exactly the properties we seek of preserving the squared angular momentum $L^2$ while reducing the energy $H$.  
This first term is briefly mentioned 
 as an example of ``double bracket dissipation'' in \cite{Bloch96}.  Conversely, when $\epsilon_H = 0$, $\bd\cdot\bL \le 0$ and $\bd\cdot\bom = 0$, so the second term preserves $H$ while reducing $L^2$.  This second term was derived purposefully in \cite{Morrison86} to produce an example of ``metriplectic'' dynamics.  The two vectors in $\bd$ point in the reciprocal directions to $\bom$ and $\bL$ in their plane, which is orthogonal to the quadratic term.  

In angular momentum space, the three principal axes consist of fixed points of the dynamics \eqref{eqofmotion} with \eqref{damping}.  In the undamped ($\bd=0$) case, the $L_1$ and $L_3$ axes consist of elliptic points and the $L_2$ axis consists of saddle points.
A linear perturbation analysis around $\bom = (\Omega, 0, 0)$, $(0, \Omega, 0)$, and $(0,0,\Omega)$, where $\Omega$ is any constant, shows that the energy-damping $\epsilon_H$ term stabilizes and destabilizes the $L_1$ and $L_3$ axes, respectively, while the momentum-damping $\epsilon_L$ term has the opposite effect.  The $L_2$ axis remains a line of saddles, and no new fixed points are created.
As in the undamped case, these terms drive the system towards collinearity of $\bL$ and $\bom$ by aligning them with a principal axis, without driving them all the way to the origin.
 The $L_1$ axis is stable if $I_1\epsilon_H > \epsilon_L$, and the $L_3$ axis is stable if $I_3\epsilon_H < \epsilon_L$.  At least one of these axes will be stable for any choice of positive coefficients, and it is possible for both to be stable simultaneously.  The final result is either a system on the $L_1$ axis with the smallest possible energy for its angular momentum, or a system on the $L_3$ axis with the smallest possible angular momentum for its energy.

\begin{figure}[h]
	\includegraphics[width=7in]{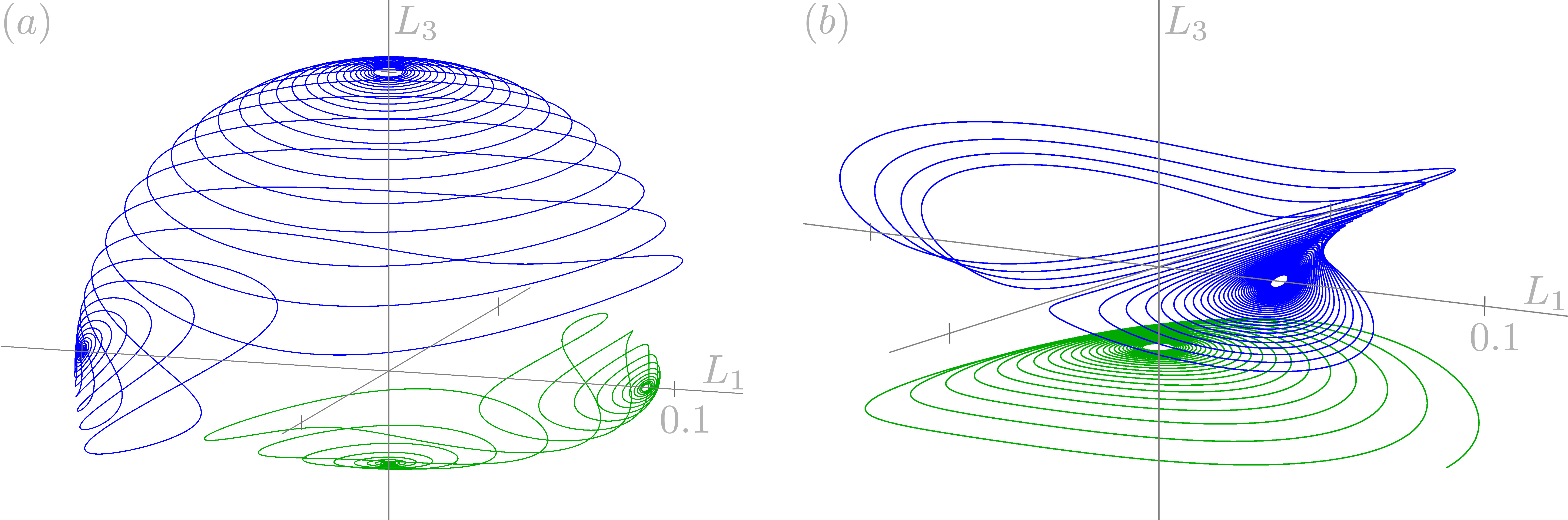}
	\caption{Example trajectories using $I_1=0.9$, $I_2=0.5$, $I_3=0.1$.  (a) Blue curve, $\epsilon_H=0.25$, $\epsilon_L=0$, only energy is dissipated and the system moves on a sphere from the $L_3$ axis to the $L_1$ axis.  Initial condition $\bom = (0.005, 0, 1)$.  Green curve, $\epsilon_H=0$, $\epsilon_L=0.25$, only angular momentum is dissipated and the system moves on an ellipsoid from the $L_1$ axis to the $L_3$ axis. Initial condition $\bom = (0.1, 0, 0.01)$.   (b) Both axes are stable when $\epsilon_H=0.25$, $\epsilon_L=0.1$. Blue curve, initial condition $\bom = (0.1, 0, 0.35)$, the system moves on a hyperboloid of one sheet, with final approach to the $L_1$ axis.  Green curve, initial condition $\bom = (0.1, 0, -0.45)$, the system moves on one half of a hyperboloid of two sheets, with final approach to the $L_3$ axis.}\label{examples}
\end{figure}

Determining which axis is reached requires a global analysis.
For any initial condition, the trajectory in angular momentum space lies on an instantaneous momentum sphere $\bL\cdot\bL = L^2$ with radius $L$ and energy ellipsoid $\bL\cdot\bom = 2H$ with largest and smallest semiaxes $\sqrt{2HI_1} \ge L$ and $\sqrt{2HI_3} \le L$. 
It can be shown that $L$ will approach or move away from these semiaxes monotonically, and the stability conditions of the linear analysis correspond to this global behavior.  
Furthermore, the attractor for a given set of initial conditions can be determined by noting that, despite the damping of both quantities, the system still preserves a first integral.
Using vector triple product identities, it is straightforward to see that 
\begin{align}
	\frac{\dot H}{\epsilon_H} = \frac{L\dot L}{\epsilon_L} = \left(\bL\cdot\bom\right)^2-\left(\bL\cdot\bL\right)\left(\bom\cdot\bom\right) = 4H^2 - L^2\omega^2 \le 0 \, ,
\end{align}
and thus 
\begin{align}
	\epsilon_H L^2 - 2\epsilon_L H = \epsilon_H \bom\cdot\bI\cdot\bI\cdot\bom - \epsilon_L\bom\cdot\bI\cdot\bom \equiv C \label{quadricvector}
\end{align}
is conserved along a trajectory.  Equation \eqref{quadricvector} defines a quadric in either angular velocity or angular momentum space.  Its representation in the latter is
\begin{align}
	\left(\epsilon_H - \frac{\epsilon_L}{I_1}\right)L_1^2 + 
	\left(\epsilon_H - \frac{\epsilon_L}{I_2}\right)L_2^2 + 
	\left(\epsilon_H - \frac{\epsilon_L}{I_3}\right)L_3^2 = C \, . \label{quadricscalar}
\end{align}
Any trajectory is confined to a quadric.  The stability transitions correspond to the change in sign of the first and third terms in \eqref{quadricscalar}, which change the system from a family of ellipsoids to a family of hyperboloids. 
Systems in which both the $L_1$ and $L_3$ axes are stable consist of hyperboloids, and changes in the sign of $C$ represent transitions between one- and two-sheeted surfaces, thus clearly selecting on which axis a particular initial condition will land.
The fixed point (pair) reached is either $L^2 = L_1^2 = \frac{I_1 C}{I_1\epsilon_H-\epsilon_L}$ or $L^2 =L_3^2 = \frac{I_3 C}{I_3\epsilon_H-\epsilon_L}$. 
A few example trajectories are shown in Figure \ref{examples}, including motion on a sphere, an ellipsoid, and hyperboloids of one and two sheets.  An explicitly computed trajectory like that on the ellipsoid can be found in \cite{MaterassiMorrison18}.

The motion of the angular momentum vector in a space-fixed frame is simply given by $\bd$. 
 In the undamped case, the vector $\bL$ is conserved in this frame, but in the general cubic case, it moves in the rotating plane defined by $\bom$ and $\bL$.  
It is worth pointing out that our approach cannot generate a dynamics wherein $\bL$, and not just $L^2$, is conserved.

\section{Discussion}

Is there any natural mechanism that could give rise to space-fixed torques of the form $\bd$ considered here?
An energy-conserving term of the form $\bom \times \bL \times \bom$ would appear as the dominant part of the corotational time derivative of $\bL \times \bom$ if the latter quantity were changing slowly in an inertial frame.  But this does not suggest any clear physical interpretation of such a term.
The other term is even more mysterious. 
On physical grounds, we might expect that the moment of inertia tensor of a deformable body could have a dependence on the angular velocities and momenta, but it does not appear that such corrections would generate terms of the form we are considering.
The two individual terms offer a means of applying torques orthogonal to the angular velocity or angular momentum, so as to conserve energy or total angular momentum, and as such are perhaps more interesting as abstract controls rather than something we might expect to see in nature.  
Quadratic control terms that conserve two quantities akin to $H$ and $L^2$ are discussed in \cite{Bloch92}, while one of the two terms considered here is discussed in \cite{MaterassiMorrison18}.  

We can construct dynamics higher order in the velocities by continuing a hierarchy of terms using the cross product.  However, this does not provide anything new, as all even and odd terms behave similarly to the quadratic and cubic terms already considered.
We have, however, neglected the possibility of higher order terms in geometric quantities that would arise from allowing the moment of inertia tensor $\bI$ to appear in a more general way.  Such terms as $\bom$, $\bI\cdot\bom$, $\bI\cdot\bI\cdot\bom$, and so on are not coplanar, so such higher order terms might provide qualitatively different dynamics.

An analogous physical system is the special case of an inextensible, anisotropic, Kirchhoff-elastic rod loaded purely by moments, so that it is force-free.  This maps onto the free rigid body by substituting an arc length derivative for the time derivative, the moment for the angular momentum, and the Darboux vector comprising the curvatures and twist of the principal material frame for the angular velocity.  The bending energy (actually the pseudo- or material momentum \cite{oreilly2017} in the absence of force) is conserved, as is the moment in the spatial frame and the magnitude of the moment in the body frame. 
Applying torques to such a rod of the form considered in this note would change these quantities accordingly, and cause an asymptotic alignment of the moment and Darboux vectors along a principal direction as one moves along the length of the rod.

A similar equation with just the energy-damping term ($\epsilon_L = 0$) can arise from the Landau-Lifshitz equation for the magnetization vector $\bM$ in a ferromagnetic material, 
$\dot \bM = \gamma \bM \times \bH - \lambda \bMh \times \bMh \times \bH$, 
where the hat denotes a unit vector and $\gamma$ and $\lambda$ are constants, if the magnetic vector field $\bH$ has a form like $\bH = \bK\cdot\bM$.  
One way this might occur is if the field is derived from an anisotropic magnetoelastic energy $-\tfrac{1}{2}\bMh\cdot\bK\cdot\bMh$ \cite{Gilbert04}, with the strains in $\bK$ that couple to the magnetization taken to be imposed constants.

%
%

Finally, it is worth asking what analogous systems might exist involving linear rather than angular velocities and momenta.  Let $\bP = \bT\cdot\bv$ be the linear momentum of a particle with velocity $\bv$ moving in an anisotropic solid that endows it with a constant effective mass tensor $\bT$.  Consider the energy $E=\frac{1}{2}\bP\cdot\bv$ and the squared magnitude of the momentum $P^2 = \bP\cdot\bP$. 
If the dynamics are given by $\dot\bP = \bd$, these quantities change as $\dot E = \bd\cdot\bv$ and $P\dot P=\bd\cdot\bP$, and again terms formed by double cross products of $\bv$ and $\bP$ will change one quantity while leaving the other unaffected.
Trajectories will lack the spiral component of the angular example, and will lie on a conic section created by the conserved quadric in linear momentum space and the conserved plane spanned by $\bP$ and $\bv$.  
Complications arise in the realignment of velocity and momentum towards hard or easy axes because the section need not contact any of the principal axes, and $\bT$ need not be positive definite.
The added mass of a body in a fluid is another more complicated situation, as in general, translational and rotational motions are coupled, with the state of the system requiring six dimensions for its description.

\section*{Acknowledgments}

I thank J. P. Domann, R. S. Hutton, and P. J. Morrison for helpful input.

\bibliographystyle{unsrt}

\end{document}